\documentclass[runningheads,a4paper]{llncs}

\usepackage{amssymb}
\usepackage{graphicx}

\usepackage{marvosym}
\usepackage{multirow}

\usepackage{algorithm}
\usepackage{algorithmic}

\usepackage{hyperref}

\usepackage{url}
\urldef{\mailsa}\path|{alfred.hofmann, ursula.barth, ingrid.haas, frank.holzwarth,|
\urldef{\mailsb}\path|anna.kramer, leonie.kunz, christine.reiss, nicole.sator,|
\urldef{\mailsc}\path|erika.siebert-cole, peter.strasser, lncs}@springer.com|

\begin{document}

\mainmatter  % start of an individual contribution

% first the title is needed
\title{An Analysis of Rank Aggregation Algorithms}

% a short form should be given in case it is too long for the running head
\titlerunning{An Analysis of Rank Aggregation Algorithms}

% the name(s) of the author(s) follow(s) next
%
% NB: Chinese authors should write their first names(s) in front of
% their surnames. This ensures that the names appear correctly in
% the running heads and the author index.
%
\author{Gattaca Lv}
\authorrunning{G. Lv}
% (feature abused for this document to repeat the title also on left hand pages)

% the affiliations are given next; don't give your e-mail address
% unless you accept that it will be published

%
% NB: a more complex sample for affiliations and the mapping to the
% corresponding authors can be found in the file "llncs.dem"
% (search for the string "\mainmatter" where a contribution starts).
% "llncs.dem" accompanies the document class "llncs.cls".
%

\toctitle{Lecture Notes in Computer Science}
\tocauthor{Authors' Instructions}
\maketitle

\begin{abstract}
Rank aggregation is an essential approach for aggregating the preferences of multiple agents. One rule of particular interest is the Kemeny rule, which maximises the number of pairwise agreements between the final ranking and the existing rankings. However, Kemeny rankings are NP-hard to compute. This has resulted in the development of various algorithms. Fortunately, NP-hardness may not reflect the difficulty of solving problems that arise in practice. As a result, we aim to demonstrate that the Kemeny consensus can be computed efficiently when aggregating different rankings in real case. In this paper, we extend a dynamic programming algorithm originally for Kemeny scores. We also provide details on the implementation of the algorithm. Finally, we present results obtained from an empirical comparison of our algorithm and two other popular algorithms based on real world and randomly generated problem instances. Experimental results show the usefulness and efficiency of the algorithm in practical settings.
\end{abstract}

\section{Introduction}

Rank aggregation has recently been proposed as a useful abstraction that has several applications in the area of both social choice theory and computer science such as meta-search, similarity search, and classification. Rank aggregation concerns how to combine many different independently constructed preferences of rankings on the same sets of alternatives by a number of different agents into a single reasonable ultimate ranking. As a result, it is intended to represent the collective opinion of the agents that constructed these rankings, and we call the collective opinion as \textquotedblleft consensus\textquotedblright. Kemeny consensus is one of the most classical and critical considerations in the rank aggregation problem for specifying a particular type of collective ranking. Given the best compromise ranking, we can sort the rankings according to their closeness to the collective one.

Unfortunately, the computational drawback of Kemeny rule is that it is known to be NP-hard in the worst case \cite{Bartholdi89,Dwork01a,DKN01}. This results in the development of various algorithms for computing Kemeny rankings. However, NP-hardness may not reflect the difficulty of solving problems that arise in practice. Therefore, in this paper we aim to compute Kemeny consensus for specific data sets such as university rankings, and thus identify which ones are closest to the consensus. More specifically, we aim to prove that Kemeny consensus can be computed efficiently during aggregating rankings in real cases when the rankings are to some extent similar with each other.

Therefore, we focus on implementing an extended version of a fixed-parameter dynamic programming algorithm \cite{Betzler09c,Betzler09a} and two other rank aggregation algorithms, which are Borda count method \cite{Borda81} and a heuristic algorithm \cite{Davenport04,Conitzer06} respectively. We also emphasise on an experimental study of the developed implementations in order to show the usefulness and efficiency of the target algorithm in practical settings. In general, the proposed solution to the above considerations is to simulate different algorithms that can take a collection of rankings on the same sets of universities as input and output corresponding Kemeny ranking and its score, together with the comparative analysis of them. For this purpose, a sample of up-to-date university rankings has been collected from real world applications. Some simulations of university rankings have also been generated randomly in order to be compared with the real cases for the evaluation.

After a series of experiments based on the models and specific data sets, we have shown that the fixed-parameter dynamic programming algorithm are capable of finding an exact optimal Kemeny solution. It also has comparable performance to the other two popular algorithms when the parameter \textquotedblleft average pairwise Kendall-Tau distance\textquotedblright\ between all university rankings is not too large.

Our paper is organised as follows. In Section 2 we describe the theory of rank aggregation and different related approaches to it. In Section 3 we specify how to implement the algorithms, in particular the dynamic programming algorithm. Then we prove the correctness of our realisations. we perform a comparative empirical study of implemented algorithms. In Section 4. Finally, in Section 5 we propose directions for future research.

\section{Background}

\subsection{Preliminaries}

Rank aggregation is a key method for aggregating the preferences of multiple agents. One rank aggregation rule of particular interest is the Kemeny rule \cite{Kem59}, which maximises the number of pairwise agreements between the final ranking and the existing rankings, and has an important interpretation as a maximum likelihood estimator. Generally, the rank aggregation problem can be described as follows.

Given a set of $m$ candidates $C = \{c_{1}, c_{2}, ..., c_{m}\}$, a ranking $\pi$ with respect to $C$ is a permutation (ordering) of all elements of $C$ which represents an agent's preference on these candidates. For each $c_{i}\in C (1\leq i\leq m)$, $\pi(c_{i})$ denotes the rank of the element $c_{i}$ in ranking $\pi$, and for any two elements $c_{i}, c_{j}\in C$, $\pi(c_{i})>\pi(c_{j})$ implies that $c_{i}$ is ranked higher than $c_{j}$ by the ranking $\pi$. In other words, we say that candidate $c_{j}$ has a greater rank than $c_{j}$ in a ranking $\pi$ if and only if $\pi(c_{i})>\pi(c_{j})$. There is a collection of $n$ rankings $\pi_{1}, \pi_{2}, ..., \pi_{n}$, which are proposed by a set of agents $A = \{1, 2, ..., n\}$ respectively. A rank aggregation method is used to get a consensus ranking $\pi$ on those $m$ candidates.

Kemeny rule is based on the concept of Kendall-Tau distance \cite{Ken38} between two rankings which counts the total number of pairs of candidates that are assigned to different relative orders in these two rankings. In other words, the Kendall-Tau distance between two rankings $\pi_{1}$ and $\pi_{2}$ is defined as:
\begin{equation}
dist_{KT}(\pi_{1},\ \pi_{2}) = |\{(c_{i}, c_{j}): \pi_{1}(c_{i})>\pi_{2}(c_{j})\hspace{4pt}and\hspace{4pt}\pi_{1}(c_{j})>\pi_{2}(c_{i})\}|
\end{equation}
Kemeny consensus is an optimal ranking $\pi$ with respect to the pre-defined $n$ rankings $\pi_{1}, \pi_{2}, ..., \pi_{n}$, which are provided by those $n$ agents and can minimise the sum of Kendall-Tau distances:
\begin{equation}
SK(\pi, \pi_{1}, \pi_{2}, ..., \pi_{n}) = \sum_{i = 1}^{n} dist_{KT}(\pi,\ \pi_{i})
\end{equation}
As we mentioned in the introductory section, the computational complexity of the problem of finding an optimal Kemeny consensus is NP-hard. $\sum_{i = 1}^{n}  dist_{KT}(\pi,\ \pi_{i})$ defined above is the score of a ranking $\pi$  with respect to the collection of rankings $\pi_{1}, \pi_{2}, ..., \pi_{n}$. Thus, the score of the optimal Kemeny consensus that minimises those sums of Kendall-Tau distances is denoted as Kemeny score.

\subsection{Rank Aggregation Algorithms}

There are numerous different rank aggregation algorithms that have been proposed in recent years. A good overview of rank aggregation methods is given in \cite{Lin10}. In general, there are two main classes of the approaches that are popular: positional approaches such as the Borda count \cite{Borda81} and majority ranking approaches such as Condorcet approaches \cite{CS05}. The Kemeny rule \cite{Kem59} is another rank aggregation rule, since it has been proposed as a way of looking for a compromise ranking. The Kemeny rule is defined as follows: it produces a ranking that maximises the number of pairwise agreements with the votes, where we have a pairwise agreement whenever the ranking agrees with one of the votes on which of a pair of candidates is ranked higher.

Greedy heuristic \cite{Cohen98} or tractable multi-stage algorithms \cite{Black89} have been developed that combine both positional and majority voting approaches. In \cite{Davenport04,Conitzer06}, computational studies for the efficient computation of a Kemeny consensus using heuristic algorithms have been performed. A new approach \cite{Dwork01a,DKN01,Didehvar07} have been proposed to produce good approximation of the optimal Kemeny consensus, and are very helpful for us to pursue approximation solutions in our context. Besides, we consider \cite{Betzler09c,Betzler09a} are really vital to our research, being an approach to take advantage of specific aspects of the data that we are hoping will be a feature of our data sets. They prove that a fixed-parameter dynamic programming algorithm could compute the Kemeny score efficiently whenever the preferences of ranking proposed by any two agents are similar with each other on average. In general, their theoretical results encourage this work for practically relevant, efficiently solvable specific data sets such as university rankings.

We describe a dynamic programming algorithm \cite{Betzler09c} originally proposed to compute Kemeny score. We extend it to find an exact optimal Kemeny consensus. The motivation of choosing this algorithm with respect to the parameter \textquotedblleft average pairwise Kendall-Tau distance\textquotedblright\ comes from some experimental studies \cite{Davenport04,Conitzer06}. It indicates that the Kemeny consensus is easier to compute when the rankings are close to each other, since we believe that the data sets of university rankings obtained from real world have such characteristics.

\begin{algorithm}[h]
\scriptsize{
\caption{The fixed-parameter dynamic programming algorithm}
\label{alg1}
\begin{algorithmic}[1]
\STATE \emph{Initialise:}
\FOR{$i=0$ to $m-1$}
\FOR{all $u\in R_{i}$}
\FOR{all $R_{i}'\subseteq R_{i}\backslash\{u\}$}
 \STATE $T(i, u, R_{i}') = +\infty$
 \ENDFOR
 \ENDFOR
 \ENDFOR
 \FOR{all $u\in R_{0}$}
 \STATE $T(0, u, \emptyset) = pK(u, U \backslash \{u\})$
 \ENDFOR
 \STATE \emph{Update:}
 \FOR{$i=0$ to $m-1$}
\FOR{all $u\in R_{i}$}
\FOR{all $R_{i}'\subseteq R_{i}\backslash\{u\}$}
\IF{$\mid R_{i}'\cup\bigcup_{j=0}^{i}F(i)\mid) = i-1$
and $T(i-1, u', (R_{i}'\cup F(i))\backslash\{u'\})$ is defined}
 \STATE $T(i, u, R_{i}') = \min_{u'\in R_{i}'\cup F(i)}T(i-1, u', (R_{i}'\cup F(i))\backslash\{u'\})+pK(u,(R_{i}\cup\bigcup_{j=i+1}^{m-1}I(j)) \backslash(R_{i}'\cup\{u\}))$ and storing $u'$
 \ENDIF
 \ENDFOR
 \ENDFOR
 \ENDFOR
 \FOR{$i=m$ to $1$}
 \IF{$T(i-1, u', (R_{i-1}'\backslash\{u'\})$ is defined}
 \STATE add $u'$ that minimises $T(i-1, u, R_{i-1}')$ to the optimal Kemeny consensus at rank $i-1$
 \ENDIF
 \ENDFOR
\STATE \emph{Output:}
\STATE the optimal Kemeny consensus and its K-score $=\min_{u\in R_{m-1}}T(m-1, u, R_{m-1}\backslash\{u\})$
\end{algorithmic}}
\end{algorithm}

Our extended dynamic programming algorithm is able to compute the exact optimal Kemeny consensus. The algorithm is depicted in Algorithm 1. The input is a collection of university rankings for $U$ given $A$, and for every $0\leq i<m$, the set $R_{i}$ of universities that can assume rank $i$ in an optimal Kemeny consensus. The output is the optimal Kemeny consensus and its Kemeny score: for every entry $T(i, u, R_{i}')$, we additionally store a university $u^{'}$ that minimises $T(i-1, u', (R_{i}'\cup F(i))\backslash\{u^{'}\})$ in line \emph{24}. Then, starting with a minimum entry for position $m-1$, we reconstruct an optimal Kemeny consensus by iteratively adding the predecessor university.

\section{Implementation of Algorithms}

The Borda count method and the heuristic algorithm not only share the same data structures but also they are much simpler to model than the dynamic programming algorithm. In this paper, we only present the realisation of the dynamic programming algorithm. For a detailed description and source code of implementations for all three algorithms, please see the author's website. Since we aim to compare different algorithms, our implementation is based on the same framework that composes of three components, which are data representation and preprocessing for representing and preprocessing obtained datasets of rankings; processing for applying an algorithm to process the representation of the preprocessed datasets; output for obtaining results such as the consensus and best aligned tables after processing.

\subsection{Data Structures}

We implement the algorithms using an objected-oriented programming language, and we specify it by a number of classes and functions. For a collection of $m$ universities, we use a $ <University>$ array $U$ to represent it, thus the size of $U$ is $m$. With regards to $n$ different agents, we use a $ <Agent>$ array $A$, thus the size of $A$ is $n$.

Computation of the partial Kemeny scores that are obtained by subdividing the overall Kemeny score is a key manipulation. We thus design an class $PrefGraph$ which acts as a intermediate data structure for storing preprocessing outcome of profile of $ulTable$. An object of $PrefGraph$ is actually a weighted directed graph, where each vertex represents one university, and the weight of each edge between vertices represents the number of agents who prefer (rank higher) university represented by start vertex to one represented by end. By this graph, it can be used to compute the partial Kemeny scores rather quickly.

To define the data structures, some further notations are needed. There are 3 sets of universities are defined: $R_{i}$, $I(i)$ and $F(i)$. For any rank $i$ from $0$ to $m-1$, $R_{i}$ denotes the set of all possible universities that can take this rank, that is, $R_{i} = \{u\in U\mid r_{ave}(u) - d < i < r_{ave}(u) + d\}$; $I(i)$ denotes the set of universities that could be \textquotedblleft inserted\textquotedblright\ at rank $i$, that is $I(i) = \{u\in U\mid u\in R_{i}\wedge u\notin R_{i-1}\}$; $F(i)$ denotes the set of universities that must be \textquotedblleft forgotten" at least at this rank, that is $F(i) = \{u\in U\mid u\in R_{i-1}\wedge u\notin R_{i}\}$. All these sets could be represented by a $<University>ArrayList$, thus we then encapsulate these $ArrayList$ into three separate classes: $RSet$, $ISet$, and $FSet$.

We use the term \textquotedblleft three dimensional dynamic programming table\textquotedblright\ to describe an abstract mechanism where we save information (the integer value) that we can later retrieve. The values depend upon the \textquotedblleft minimum partial Kemeny score\textquotedblright\ over all possible orders of the universities of $R_{i}'$ given u taking rank $i$ and all universities of $R_{i}'$ taking ranks below $i$. Thus, it is represented by a three-dimensional array $T$, where the first dimension is rank $i$, the second dimension is every university $u$ can assume $i$, and the third dimension is every university subset $R_{i}'\subseteq R_{i}\backslash\{u\}$.

\subsection{Critical Functions}

\subsubsection{generatePrefGraph()}

is used to generate a preference graph given a collection of university rankings. An object of the abstract data type $PrefGraph$ is a directed weighted graph, and it is represented by an adjacency matrix in the implementation. We use a simple example to describe the preference graph and its representation, and they are also showed in Fig. 1.

\begin{figure}[h!]
\begin{minipage}[c]{0.5\linewidth}
\centering
\includegraphics[height=3cm]{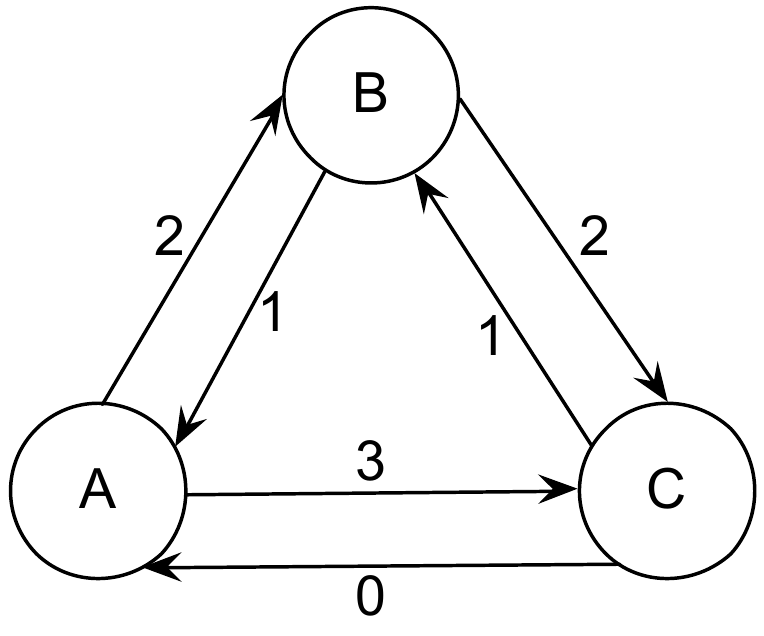}
\end{minipage}
%\hfill
\begin{minipage}[c]{0.2\linewidth}
\centering
\includegraphics[height=3cm]{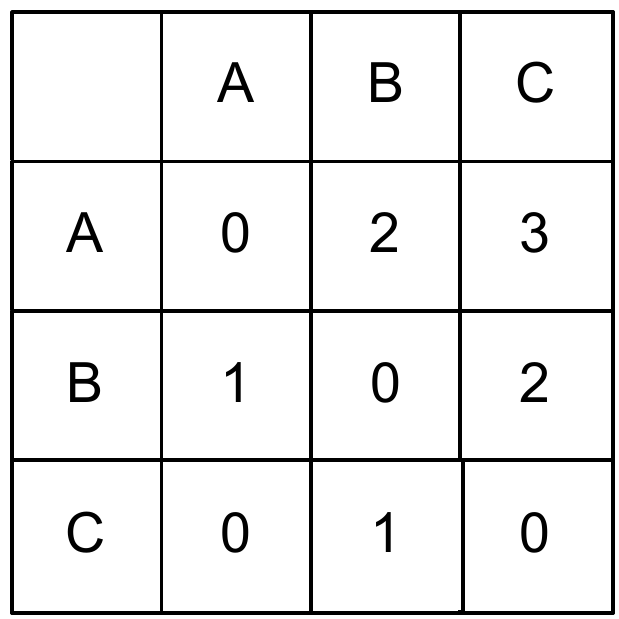}
\end{minipage}
\caption{Preference graph and its adjacency matrix representation.}
\end{figure}

Considering three university rankings expressed by the profile ((0, $A$), (1, $B$), (2, $C$)), ((0, $A$), (1, $C$), (2, $B$)), ((0, $B$), (1, $A$), (2, $C$)), this means that: the first agent ranks university A better than university B than university C, the second agent ranks A better than C than B, the third agent ranks B better than A than C. The preferences for each of the pairwise rankings are: $a_{AB}=2,a_{BA}=1,a_{BC}=2, a_{CB}=1, a_{AC}=3, a_{CA}=0$.

\subsubsection{computeAveRank()}

is used to return the average rank of each university given a collection of university rankings. Let the rank of a university $u$ in a university ranking proposed by a gent $a$, denoted by $r_{a}(u)$, be the number of universities that are better than $u$ in $a$. That is, the topmost and best university in $a$ has rank 0 and the bottommost has rank $m-1$. For a collection of university rankings for $U$ given $A$ and university $u\in U$, the average rank $r_{a}(u)$ of $u$ is defined as:
\begin{equation}
r_{ave}(u) = \frac{1}{n}\cdot\sum_{a\in A}^{}  r_{a}(u)
\end{equation}
The average ranks of all universities can be computed in $O(n\cdot m)$ time by iterating once over every university ranking and adding the rank (an integer value) of every university to a counter variable for this university.

\subsubsection{computeAveKTdistance()}

is used to return the average pairwise Kendall-Tau distance between a given collection of university rankings. For a collection of university rankings for $U$ given $A$, the average Kendall-Tau distance $d_{ave}(u)$ is defined as:
\begin{equation}
d_{ave} = \frac{2}{n(n-1)}\cdot\sum_{a,a'\in A,a\neq a'}^{} dist_{KT}(a,a')
\end{equation}
where for each pair of university ranking $a$ and $a'$, the Kendall-Tau distance (KT-dist) between universities $u$ and $u'$ is defined as:
\begin{equation}
dist_{KT}(a,a') = \sum_{{u,u'}\subseteq U}^{} d_{a,a'}(u,u')
\end{equation}
where the sum is taken over all unordered pairs ${u,u'}$ of universities, and $d_{a,a'}(u,u')$ is 0 if $a$ and $a'$ rank $u$ and $u'$ in the same order, and 1 otherwise.

Furthermore, the value of $d_{ave}$ may be a non-integer. However, for the purpose of the whole dynamic programming algorithm, this method is supposed to return a $<int>$ value, thus $d$ is defined as: $d = \lceil d_{ave}\rceil$

\subsubsection{computePKScore()}

is used to compute partial Kemeny score given an university, a subset of universities excluded this university, and a collection of university rankings. As for the dynamic programming, it is necessary to subdivide the overall Kemeny score into partial Kemeny scores. More precisely, for an university $u$ and a subset $R$ of universities with $u\notin R$, we define:
\begin{equation}
pK(u,R) = \sum_{u'\in R}\sum_{a\in A}^{}  d^R_{a}(u,u')
\end{equation}
where for $u\notin R$ and $u'\in R$ we have $d^R_{a}(u,u') = 0$ if in the university ranking proposed by $a$ we have $u > u'$, and $d^R_{a}(u,u') = 1$, otherwise.

Intuitively, the partial Kemeny score denotes the score that is induced by university $u$ and the university subset $R$ if the universities of $R$ have greater ranks than $u$ in an optimal Kemeny consensus.

\subsubsection{findBestTable()}

is used to return which university rankings are closest to the Kemeny consensus given a collection of university rankings and their Kemeny consensus. The score of a university ranking $t$ with respect to a collection of university rankings for $U$ given $A$ is denoted as:
\begin{equation}
SK(t,a_{1},a_{2},...,a_{n}) = \sum_{a\in A}^{} dist_{KT}(t,a)
\end{equation}
We already know that a university ranking $t$ with the minimum score is called an optimal Kemeny consensus of $(U,A)$, thus we aim to identify which university rankings given by agents are closet to the consensus.

\subsection{Correctness of the Implementation}

We test all possible universities for every rank $0\leq i \leq length-1$. Thus, having chosen an university $u$ for rank $i$, the remaining universities that could assume $i$ must either has smaller rank or bigger than i in an optimal Kemeny consensus. To ensure the correctness of the implementation, we show that it satisfies the following two conditions:

First, the capability of the realisation to find an optimal Kemeny solution is ensured. We know that the Kemeny score can be decomposed into partial Kemeny scores, thus we can show that the algorithm considers a decomposition that results in the final optimal Kemeny consensus. For every rank, the algorithm tests every university in $R_{i}$. Based on the definition of the set $R_{i}$ of universities for rank $i$, one  of these universities must be the correct university $u$ for this rank. Furthermore, for $u$ we are able to find that the algorithm tests a sufficient number of possibilities to partition all remaining universities $U \backslash \{u\}$ such that they either be left or right of rank $i$. More precisely, every university from $U \backslash \{u\}$ must be in exactly one of the following three subsets:

\begin{itemize}
\item The set $F$ of universities that have already been forgotten, that is, $F=\bigcup_{j=0}^{i}F(j)$.
\item  The set of universities that can assume rank $i$, that is, $R_{i}\backslash\{u\}$.
\item  The set $I$ of universities that are not inserted yet, that is, $I= \bigcup_{j=i+1}^{m-1}I(j)$.
\end{itemize}

Second, all entries in the three dimensional table are well defined and its value is computed correctly. For any entry $T(i, u, R_{i}') $, in terms of rank $i$ there must be exactly $i-1$ number of universities that have ranks smaller than $i$.

%--------------------------------------------------------------------------------------------

\section{Empirical Evaluation}

In this section, we present a critical appreciation of the strengths and limitations of our implementations. To the best of knowledge, there is no existing empirical comparison of different algorithms. As a novelty, we aim to compare the actual effectiveness and efficiency of these algorithms in real case.

\subsection{Data Sets}

We will use three kinds of data sets of university rankings obtained from both reality and simulation, which are denoted by,  $SAME$, $DIFF$, and $RANDOM$ respectively. First, $SAME$ provides five university rankings, and they are obtained from some consecutive years of rankings that were published by the \textquotedblleft same\textquotedblright\ organisation in real case. Second, $DIFF$ offers the same year rankings proposed by some \textquotedblleft different\textquotedblright\ organisations. Third, $RANDOM$ denotes the data sets of five university rankings, in which all universities and their ranks are generated \textquotedblleft randomly\textquotedblright\ during the runtime of each execution in the experiments. For the convenience of our analysis, we have chosen a set of the same 40 universities. Thus, the full list in each university ranking consists of exactly the same 40 universities in the $SAME$, $DIFF$, and $RANDOM$ data sets. Further, we evaluate different rank aggregation algorithms in terms of their usefulness and efficiency. We consider subsets of the data sets, that are, $SAME_{m}$, $DIFF_{m}$, and $RANDOM_{m}$, where $m$ is the length of each partial ranking.

\subsection{Experimental Setup}

After collecting the data sets, each algorithm has been evaluated in terms of its effectiveness, i.e., its average Kemeny score of the Kemeny solution was computed, and its efficiency, i.e., its average running time was also computed. We consider the execution results of $SAME_{m}$, $DIFF_{m}$, and $RANDOM_{m}$. Our methodology to examining the effectiveness and efficiency of different algorithm is as follows. Each experiment is applied to the partial data set $SAME_{m}$, $DIFF_{m}$, and $RANDOM_{m}$ individually. For each data set, each data point is the average of 20 trials. Each trial is performed as follows: for every possible $m\in\{6, 8, 10, 12\}$, randomly select a subset $U$ that contains the number $m$ universities from 40 universities in the full set, and then to form a new university ranking made up of this subset of same $m$ of universities. Therefore, for $m=6$, we have:  $SAME_{6}$, $DIFF_{6}$, and $RANDOM_{6}$, and for $m=8$, we have:  $SAME_{8}$, $DIFF_{8}$, and $RANDOM_{8}$, and so on.

\subsection{Preliminary Results}

There are four essential aspects that we aim to investigate in the experiment, which are the ranking length, average pairwise Kendall-Tau distance, Kemeny score of consensus ranking, and experimental computational complexity. Therefore, for partial data sets with different ranking length, we investigate three parameters (outputs) from three perspectives of measurement criteria, that are, minimum value, maximum value, and average value. For detailed results, please refer to the author's website as well.

Since the dynamic programming algorithm can output the exact Kemeny consensus, its effectiveness for finding optimal Kemeny ranking should be the best, and its precision is 100\%. In Fig. 2(a), we can see that the heuristic algorithm achieves a very good approximation of the exact solution, because its Kemeny score is really close to the dynamic programming algorithm. The largest Kemeny score indicates the least precision of result consensus, thus the Borda count method has the worst effectiveness among all three algorithms.

\begin{figure}[h!]
\begin{minipage}[c]{0.5\linewidth}
\centering
\includegraphics[height=4cm]{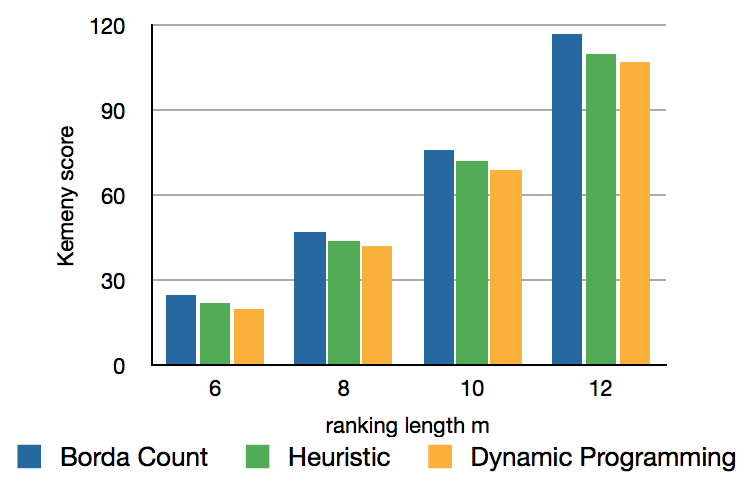}
\end{minipage}
%\hfill
\begin{minipage}[c]{0.1\linewidth}
\centering
\includegraphics[height=4cm]{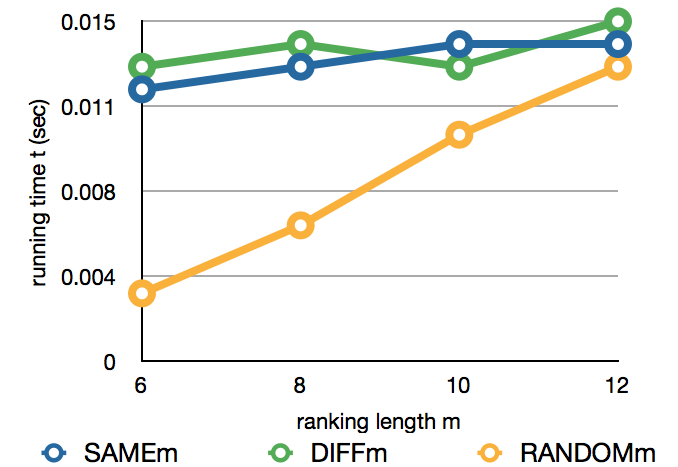}
\end{minipage}
\caption{(a) Comparison on different algorithms according to average Kemeny score. (b) Average running time of the Borda count method.}
\end{figure}

In Fig. 2(b) and 3(a) , it is not hard to see that the total running time $t$ of both Borda count method and heuristic algorithm is almost linear to the ranking length $m$ under the experiments on three different kinds of data sets of university rankings $SAME_{m}$, $DIFF_{m}$, and $RANDOM_{m}$. It indicates that the experimental computational complexity for both algorithms is constant no matter how many universities each university ranking consists of. Thus, we could test much larger $m$ (e.g., 50, 100, etc.), although that $m$ cannot be tested on dynamic programming algorithm due to its memory consumption problem in dynamic programming table construction.

In Fig. 3(b), the total running time $t$ on data set $RANDOM_{m}$ is increasing significantly if $m$ increases from 6 to 12. Because we know that $RANDOM_{m}$ has average pairwise Kendall-Tau distance $d$ much larger than $SAME_{m}$ and $DIFF_{m}$, it indicates that its time complexity is exponential to its ranking length $m$. Further, to see whether this NP-hardness could be overcame in some circumstances, we look into the $SAME_{m}$ and $DIFF_{m}$ cases applied to dynamic programming algorithm. As we can see, the performance of $SAME_{m}$ is much better than $DIFF_{m}$, and it has also clearly decreased the running time than $RANDOM_{m}$.

\begin{figure}[h!]
\begin{minipage}[c]{0.5\linewidth}
\centering
\includegraphics[height=4cm]{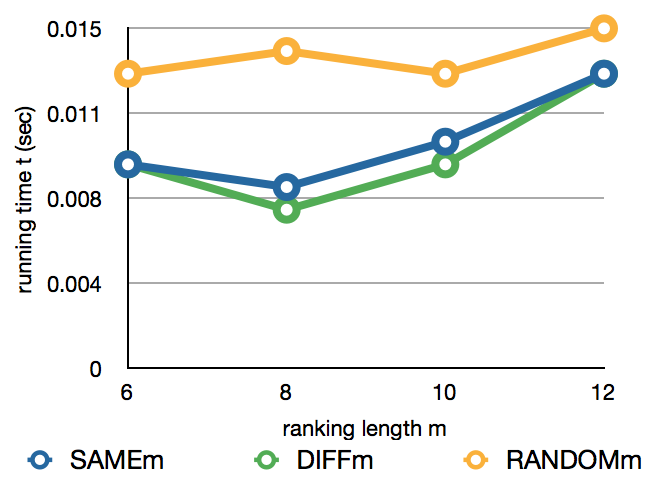}
\end{minipage}
%\hfill
\begin{minipage}[c]{0.1\linewidth}
\centering
\includegraphics[height=4cm]{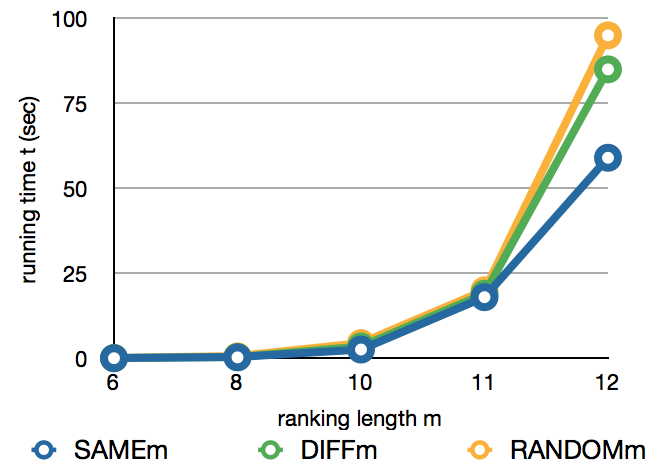}
\end{minipage}
\caption{(a) Average running time of the heuristic algorithm. (b) Average running time of the dynamic programming algorithm.}
\end{figure}

As a result, our experimental results proved that finding Kemeny consensus is fixed-parameter tractable with respect to the parameter \textquotedblleft average pairwise Kendall-Tau distance $d$\textquotedblright\ ($d_{SAME_{m}} < d_{DIFF_{m}} < d_{RANDOM_{m}}$). Further, it is easy to see that to a great extent the efficiency of dynamic programming algorithm also relies on the ranking length. To conclude, we say computing optimal Kemeny consensus can work effectively and efficiently for data set $SAME$ if there is not too many controversial universities in all rankings. Further, we still need to improve the performance of the algorithm to make it be more realistic on data set $DIFF$.

\subsection{Discussion}

There are theoretical and practical advantages with respect to the algorithms. The theoretical advantage  is that: if the average pairwise Kendall-Tau distance $d$ between a number of $m$ university rankings is relatively not too large (e.g. say: $d \leq m/2$), which means university rankings are similar each other, we could then overcome the NP-hard computational complexity and compute the Kemeny ranking effectively and efficiently no matter how many university are there in a ranking. The practical advantage is that: given a profile of a number $n$ of university rankings and each one containing a full list of the same $C$ universities, we could choose any integer number $m$ that is not larger than $C$, that is $m \leq C$, and then select any $m$ number of universities among the full list of $C$ universities in order to aggregate the new partial university rankings formed by only those $m$ universities.

Unfortunately, if $d$ is too large, the developed algorithm will be inefficient. Since the set $R_{i}$ of universities that can take rank $i$ probably contains all universities in the ranking, this will increase the workload of data processing and operation significantly. However, we could still compute the exact Kemeny optimal solution efficiently. The solution may be improving the computation power or changing the computation environment from single computing to distributed or parallel computing.

In addition, whenever $d \geq m$, and $m$ is too large at the same time, the developed program will terminate its running unexpectedly. There are $m$ ranks (universities) in a partial ranking, and any rank can be assumed by at most $4d$ universities. The number of considered subsets is bounded from above by $2^{4d}$. Hence, the size of the table $T$ is $O(2^{4d}\cdot d\cdot m)$. However, if $d \geq m$, the size of the table $T$ then become $O(2^{m}\cdot m^{2})$, which is extremely large. For instance, the computer we used has a 2GB of physical memory. When $m = 21$ and $d \geq 21$, it will need $2^{21}\cdot 21^{2}\cdot2 = 2.0\times10^{9}$ Bytes = 2GB of space for storing the dynamic programming table, which necessarily exceeds the available memory of the machine.

\section{Future Work}

There are several possible directions for future work. First, we can extend our cases such that university rankings may have ties or are incomplete. As for the number of rankings, if the number is even, there may be a tie between these ranking \cite{FKM+04,Ken45}. It is also worth to study the case that a university in a ranking may not appear in another one. Second, we can extend our cases to rankings that may have weights. We measure that different ranking may have different implication, so we assign a real number to it as the factor in its importance. Third, we can improve the running time as well as the memory consumption of the three-dimensional dynamic programming table. Finally, we can implement other fixed-parameter algorithms with respect to other parameters such as \textquotedblleft maximum range\textquotedblright\ and \textquotedblleft average range\textquotedblright\ of ranks \cite{Betzler09c,BFG09b,Betzler09a}.

\bibliographystyle{splncs}
\bibliography{mylncs}

\begin{thebibliography}{10}

\bibitem{Bartholdi89}
Bartholdi, J.J., Tovey, C.A., Trick, M.A.:
\newblock Voting schemes for which it can be difficult to tell who won the
  election.
\newblock Social Choice and Welfare \textbf{6}(2) (1989)  157--165

\bibitem{Dwork01a}
Dwork, C., Kumar, R., Naor, M., Sivakumar, D.:
\newblock Rank aggregation methods for the web.
\newblock In: proceedings of the 10th international conference on World Wide
  Web, Hong Kong, ACM (2001)  613--622

\bibitem{DKN01}
Dwork, C., Kumar, R., Sivakumar, D., Naor, M.:
\newblock Rank aggregation revisited.
\newblock Technical report, Compaq Systems Research Center (2001)

\bibitem{Betzler09c}
Betzler, N., Fellows, M., Guo, J., Niedermeier, R., Rosamond, F.:
\newblock Fixed-parameter algorithms for computing kemeny scores.
\newblock In: proceedings of The 4th International Conference on Algorithmic
  Aspects in Information and Management. (2008)  60--71

\bibitem{Betzler09a}
Betzler, N., Fellows, M.R., Guo, J., Niedermeier, R., Rosamondb, F.A.:
\newblock Fixed-parameter algorithms for kemeny rankings.
\newblock Theoretical Computer Science \textbf{410}(45) (2009)  4554--4570

\bibitem{Borda81}
de~Borda, J.C.:
\newblock M{\'e}moire sur les {\'e}lections au scrutin.
\newblock In: Histoire de l'Academie Royal des Sciences, Paris (1781)

\bibitem{Davenport04}
Davenport, A., Kalagnanam, J.:
\newblock A computional study of the kemeny rule for preference aggregation.
\newblock In: proceedings of the 19th national conference on Artifical
  intelligence, AAAI Press (2004)  697--702

\bibitem{Conitzer06}
Conitzer, V., Davenport, A., Kalagnanam, J.:
\newblock Improved bounds for computing kemeny rankings.
\newblock In: proceedings of the 21st national conference on Artificial
  intelligence. (2006)  620--626

\bibitem{Kem59}
Kemeny, J.:
\newblock Mathematics without numbers.
\newblock Daedalus \textbf{88}(4) (1959)  577--591

\bibitem{Ken38}
Kendall, M.:
\newblock A new measure of rank correlation.
\newblock Biometrika \textbf{30}(1/2) (1938)  81--93

\bibitem{Lin10}
Lin, S.:
\newblock Rank aggregation methods.
\newblock Wiley Interdisciplinary Reviews: Computational Statistics
  \textbf{2}(5) (2010)  555--570

\bibitem{CS05}
Conitzer, V., Sandholm, T.:
\newblock Common voting rules as maximum likelihood estimators.
\newblock In: In Proceedings of the 20th Conference on Uncertainty in
  Artificial Intelligence (UAI2005), Morgan Kaufmann Publishers (2005)
  145--152

\bibitem{Cohen98}
Cohen, W.W., Schapire, R.E., Singer, Y.:
\newblock Learning to order things.
\newblock In: Proceedings of the 1997 conference on Advances in neural
  information processing systems, MIT Press (1998)  451--457

\bibitem{Black89}
Black, D.:
\newblock The theory of committees and elections.
\newblock European Journal of Political Economy \textbf{5}(4) (1989)  595--596

\bibitem{Didehvar07}
Didehvar, F., Eslahchi, C.:
\newblock An algorithm for rank aggregation problem.
\newblock Applied Mathematics and Computation \textbf{189}(2) (2007)
  1847--1858

\bibitem{FKM+04}
Fagin, R., Kumar, R., Mahdian, M., Sivakumar, D., Vee, E.:
\newblock Comparing and aggregating rankings with ties.
\newblock In: Proceedings of the 23rd ACM SIGMOD-SIGACT-SIGART symposium on
  Principles of database systems (PODS '04), ACM (2004)  47--58

\bibitem{Ken45}
Kendall, M.G.:
\newblock The treatment of ties in ranking problems.
\newblock Biometrika \textbf{33}(3) (1945)  239--251

\bibitem{BFG09b}
Betzler, N., Fellows, M., Guo, J., Niedermeier, R., Rosamond, F.A.:
\newblock How similarity helps to efficiently compute kemeny rankings.
\newblock In: proceedings of The 8th International Conference on Autonomous
  Agents and Multiagent Systems. (2009)  657--664

\end{thebibliography}

\end{document}